\documentclass[final,5p,times,twocolumn,sort&compress]{elsarticle}
\usepackage[author={DLG}]{pdfcomment}

\usepackage{xcolor}
\usepackage{graphicx}

\usepackage[english]{babel}

\usepackage{mathtools}
\usepackage{bm}
\usepackage{bbm}

\begin{document}
\frontmatter

\title{Clipped DeepControl: deep neural network two-dimensional pulse design with an amplitude constraint layer}

\author[CFIN]{Mads Sloth Vinding\corref{cor1}}
\ead{msv@cfin.au.dk}
\author[CFIN]{Torben Ellegaard Lund}

\cortext[cor1]{Corresponding author.}
\address[CFIN]{Center of Functionally Integrative Neuroscience (CFIN), Department of Clinical Medicine, Faculty of Health, Aarhus University, Denmark}
\date{\today}
\begin{abstract}
Advanced radio-frequency pulse design used in magnetic resonance imaging has recently been demonstrated with deep learning of (convolutional) neural networks and reinforcement learning. 
For two-dimensionally selective radio-frequency pulses, the (convolutional) neural network pulse prediction time (few milliseconds) was in comparison more than three orders of magnitude faster than the conventional optimal control computation. 
The network pulses were from the supervised training capable of compensating scan-subject dependent inhomogeneities of $B_0$ and $B^+_1$ fields. 
Unfortunately, the network presented with a non-negligible percentage of pulse amplitude overshoots in the test subset, despite the optimal control pulses used in training were fully constrained. 
Here, we have extended the convolutional neural network with a custom-made clipping layer that completely eliminates the risk of pulse amplitude overshoots, while preserving the ability to compensate the inhomogeneous field conditions. 
\end{abstract}

\maketitle 
\section{Introduction}
In magnetic resonance imaging (MRI), the subject to be imaged is placed in a (main) magnetic field, denoted $B_0$, nominally and typically of 1.5, 3, or 7 Tesla (T) formed by a large, superconducting solenoid. 
Hereby, certain nuclear-magnetic-resonance active atomic nuclei (e.g. of Hydrogen) form a small magnetization, which can be probed with the corresponding (Larmor) resonance frequencies of 64, 128, and 300 MHz, respectively. 
By a coil nearby, a second perturbing electromagnetic field, denoted $B^+_1$, transmits radio-frequency (RF) pulses into the subject, e.g., to invert, excite or refocus the magnetization, as a mean to efficiently exploit the magnetization's ability to induce a measurable electrical current in a receive coil nearby, which has an associated receive field denoted $B^-_1$. 
Of essence particularly for MRI, pulsed magnetic field gradients are used to encode spatial information into the signal. 
Magnetic field gradients stem from pulsed currents running typically in two Golay saddle-coils and one circular Maxwell coil, covering the $x$-, $y$- and $z$-directions, respectively. 

MRI has a plethora of different RF and gradient pulse schemes for different imaging applications. 
Simple fixed-shape RF pulses are typically one-dimensional (e.g., a Hamming filtered sinc-shape) and therefore just selects the image slice (a slice of the imaged body)\cite{brown_magnetic_2014}. 
Neither the $B_0$, $B^+_1$, or $B^-_1$ fields are perfectly homogenous, and they depend partly on the subject being scanned. 
Variations are expected and may be seen in images as distortions, blurring, shading etc. 
This communication concerns RF pulse design, where $B_0$ and $B^+_1$ fields are part of the underlying math. 
The $B^-_1$ field inhomogeneity and compensation thereof is another matter that will not be discussed further, but one account of this is for example the work of Marques et al.~\cite{marques_mp2rage_2010}. 
Simple RF pulses do not compensate spatial $B_0$ and $B^+_1$ field inhomogeneities, or at least not to the same extent as more advanced pulse designs.
Advanced pulses may be selective in all three spatial dimensions as well as the spectral dimension if needed. 
For instance, 2D RF pulses are not slice selective, but selective in the 2D image plane itself, and are therefore able to encode and compensate $B_0$ and $B^+_1$ field inhomogeneities in the image plane\cite{schoenberg_parallel-excitation_2007}. 
They may, e.g., be used to zoom the selection to a small region of interest, which the following image acquisition can focus on, resulting, e.g., in faster scanning with less energy deposition for the same resolution\cite{finsterbusch_fast-spin-echo_2010}. 
Yet, to compensate subject-dependent field inhomogeneities, these tailored pulses need maps of these $B_0$ and $B^+_1$ fields as they need to plug into the math. 
While these maps can be measured rather swiftly nowadays\cite{herrler_fast_2021}, exploiting the field maps in tailored pulse designs typically require lengthy iterative numerical optimizations. 
In fact, the urge to achieve more difficult applications, e.g., multi-dimensional, multi-channel transmit at ultrahigh field (UHF), while mitigating more and more demanding experimental challenges such as increasing $B^+_1$ inhomogeneities at UHF\cite{schoenberg_parallel-excitation_2007}, or complying to more and more demanding constraints, e.g., pulse power and local specific absorption rate (SAR), the pulse design methods often become too slow in practice and hamper clinical applicability\cite{vinding_local_2017}.

\begin{figure*}[t!]
\includegraphics[width=\textwidth, angle=0]{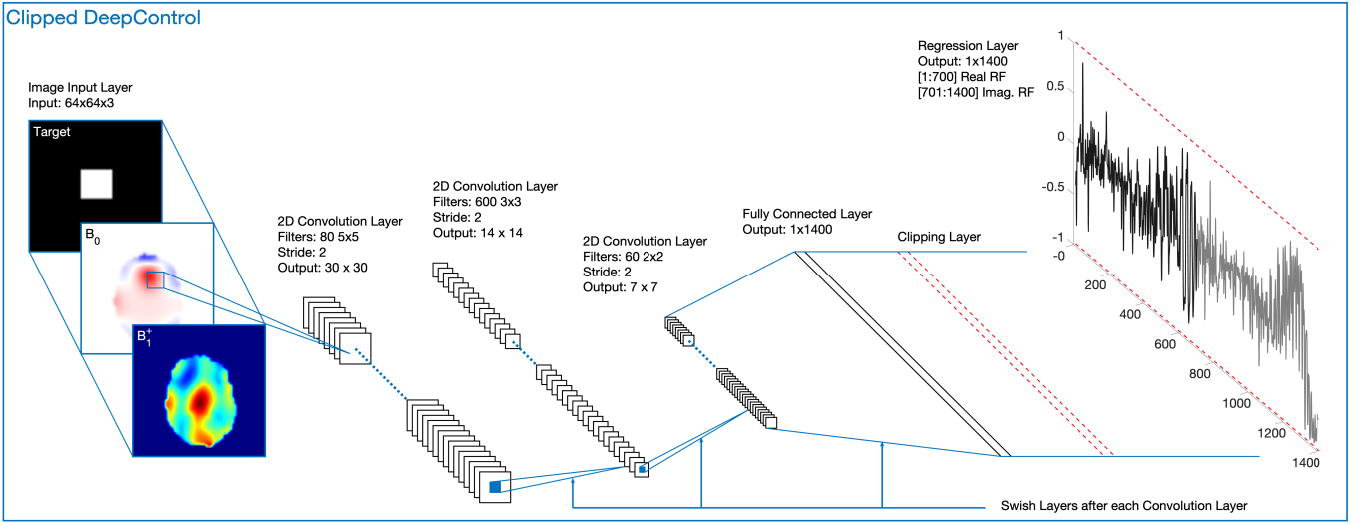}
\caption{The DeepControl method including the clipping layer. From the left, the image input layer loads a stack of three images of a given size into the network, mapping first of all the 2D RF pulse target area (here as a binary mask). Secondly, the $B_0$ field map showing the main magnetic field inhomogeneity. Thirdly, the RF coil's sensitivity through a $B^+_1$ field map. These maps are among three typical input parameters in 2D RF pulse design. The hidden layers perform 2D convolution followed by Swish layer activation functions. Before the final regression layer there is a fully connected layer that matches the desired output array, and as the central theme in this communication a custom-made two-sided clipping layer that prevents the regression layer output from exceeding a user-defined limit by simple cutting. The regression layer contains the real and imaginary 2D RF pulse in a concatenated array. When using the network predicted pulse in a simulation or experiment, it must be accompanied with the given magnetic field gradient waveform shown in Figure~\ref{FigPulses}. This magnetic field gradient waveform was used, when building the supervised training library with the conventional 2D RF pulse optimization framework.}
\label{FigMethodDeepControl}
\end{figure*}

In order to circumvent the long computations, to approach a real-time push-button solution, we recently introduced a deep learning neural network method\cite{vinding_ultrafast_2019}.
It used supervised learning of a vast library of 2D RF pulses made from different numerical optimization approaches, e.g., the optimal control methods, we have reported earlier\cite{vinding_local_2017,vinding_optimal_2021-1,maximov_real-time_2015}. 
Firstly, we showed a fully connected neural network trained upon a few thousand training examples could mimic the optimal control optimizations. 
Importantly, the neural network pulse prediction times were on average 7 ms, which was more than three orders of magnitude faster than the optimal control computations\cite{vinding_ultrafast_2019}. 
In the follow-up, we extended the neural network with convolution layers in order to better facilitate $B_0$ and $B^+_1$ field maps, and we demonstrated compensation of these field inhomogeneities in UHF experiments\cite{vinding_deepcontrol_2021}. 

In both studies, however, the RF pulse amplitudes were constrained only in the optimal control training library. 
Meaning that we relied on a fully constrained training library with respect to pulse power, and the network's ability to deep learn those limits. 
I.e., we had no handle on pulse power in the network pulse predictions. 
Overall, the statistical risk of pulse amplitude overshoot was benign, e.g., 2\% of the test cases were overshooting in the last study, and the amount of overshoot was not alarming, and we described methods to handle overshoots\cite{vinding_deepcontrol_2021}. 
However, as we have later developed networks for RF pulses with stronger excitation flip angles of the magnetization (going from 30$^\circ$ in Ref.~\cite{vinding_deepcontrol_2021} to the 90$^\circ$ herein), we observed an unsatisfying greater risk of pulse amplitude overshoot. 

As we suggested in Ref.~\cite{vinding_deepcontrol_2021}, this communication reports on a convolutional neural network that includes a custom-made clipping activation function, inserted as a layer, which completely prevents pulse amplitude overshoots without loss in performance and with the preserved ability to compensate inhomogeneous field conditions. 
\section{Method}
The method presented here is an updated version of that of Ref.~\cite{vinding_deepcontrol_2021}. 
In short, a vast supervision library is established, see Section~\ref{sec:library}, where each entry has an input and an output. 
The input mimics a pulse design situation through three images: a 2D RF target pattern, a $B_0$ map and a $B^+_1$ maps. 
The output for the same library entry contains the 2D RF pulse that, based on the input, is computed by a conventional pulse design method, e.g., optimal control (Section~\ref{sec:oc}). 
The convolutional neural network, Section~\ref{sec:cnn}, was designed first of all to handle the specific input and target array sizes, plus the novel clipping layer was included (Section~\ref{sec:clip}). 
The hidden layers, activation functions etc. were adjusted by monitoring the deep learning progress during several trial runs.
All computations were performed in MATLAB R2021a (Mathworks, Natick, MA, USA) on a $28\times 2.2$ GHz Intel Xeon Gold 5120 (Intel Corporation, Santa Clara, CA) workstation with 384 GB RAM.
\subsection{Supervision library}
\label{sec:library}
The total library consisted of five hundred thousand individual entries. 
The input part of each entry consists of three maps of 64-by-64 pixels.
An input entry is shown in Figure~\ref{FigMethodDeepControl} (left).
Here we show a realistic case to illustrate the DeepControl method in operation.
The first maps out (spatially) the target region of interest as a binary mask, i.e., what the 2D RF pulse is aimed to select for zoomed imaging. 

The other two are the $B_0$ and $B^+_1$ maps. 
In the library, however, the three images are not MRI related as such.
The target patterns are made by placing ten points randomly on the grid as corners in a polygon. 
This polygon is then morphologically dilated and eroded twice with disk shaped structure elements of radius 8, 12, 3 and 4 pixels, respectively.

As seeds for $B_0$ and $B^+_1$ field maps, we used photographs from ImageNet\cite{j_deng_imagenet_2009} cut to 64-by-64 pixels. 
This choice was made to induce a large variability in map patterns that with respect to typical MR images would seem random. 
The $B_0$ photograph seeds were firstly subtracted by their median pixel value and then normalized. 
The normalized map was then multiplied by a random value selected from the ranges $[-600\,\mathrm{Hz},-50\,\mathrm{Hz}]$ and $[50\,\mathrm{Hz},600\,\mathrm{Hz}]$.
These values correspond to typical maximum offset values observed in measured $B_0$ maps.
The relation to frequency units is given by the Larmor precession frequency equation $f_0 = \frac{\gamma}{2\pi} B_0 $, where $\gamma$ is the natural gyromagnetic ratio constant for the given nucleus (here the $^1\mathrm{H}$ nucleus).
The excluded range from $(-50\,\mathrm{Hz},50\,\mathrm{Hz})$ assured that the $B_0$ maps were not too benign.

A $B^+_1$ field rotates the magnetization by an amount denoted a flip angle (FA). 
And as such, the target map displays a region with a desired FA.
In this study, a value of 0 or 1 in the target map correspond to a desired, nominal FA of $0^\circ$ or $90^\circ$, respectively. 
The actually achieved FA (depending on $B^+_1$), though, is found by experiment or some form of RF pulse simulation, e.g., the generally applicable Bloch equations, or for small FAs the Fourier transform. 
An inhomogenous $B^+_1$ may therefor not achieve the nominal FA everywhere, if not attended to in the RF pulse design.
The $B^+_1$ photograph seeds were just normalized to the internal between $[0,1]$. 
A value of 1 in such a sensitivity map corresponds to a point, where the $B^+_1$ field is capable of achieving the nominal FA. 

A common superelipsis mask was further applied to the maps to simulate potential object boundaries, with the outside area working as a "don't-care" region that is not included in the conventional pulse optimizations either.
Because the target pattern had values of 0 or 1, and $B^+_1$ maps had values also in between 0 and 1, the $B_0$ maps were furthermore downscaled by 600 Hz to achieve values between -1 and 1 (depending on the random generator). 
This was done to avoid mixing different dynamic ranges in the library that could lead to numerical instability, 

\subsubsection{Optimal control}
\label{sec:oc}
As a supervision library the input maps described in the previous section need an output part as well. 
This is in our studies 2D RF pulses computed with optimal control. 
The optimal control applies, through the Bloch equations, all physically relevant parameters, which the convolutional neural network is blind to. 
The target maps are translated to a nominal FA of $90^\circ$, and the $B_0$ field maps are rescaled back by 600 Hz, when drawn from the library.

The $B^+_1$ field maps are plugged in as sensitivity maps and stay unit-less. 
The 2D RF pulses will then be computed in field strength or frequency units by the relation $f_1 = \tfrac{\omega_1}{2\pi} = \tfrac{\gamma B_1}{2\pi}$. 
The frequency unit refers to the \textit{rate} by which the magnetization is rotated as a result of the applied $B^+_1$ field.
In turn, we can impose limits on the 2D RF pulse amplitudes (real and imaginary channels) in frequency units. 
Our RF amplitude constraint was $c' = 1\,\mathrm{kHz}$, meaning the real and imaginary channels could vary freely in the range $[-c',+c']$.
The choice of using frequency units etc. is one common convention.
Among other conventions, it is also common to measure $B^+_1$ field maps in Tesla per Volt or Tesla per Watt, which directly refers to the electric potentials or powers applied to the RF coils by the RF amplifiers, respectively.
RF pulses are then output and constrained in other units like Volt or Watt\cite{vinding_local_2017}. 

The maps of 64-by-64 pixels are setup to span a 25-cm square. 
This is done by the underlying magnetic field gradient waveform, shown later, that encodes spatial information. 
The magnetic field gradient waveform dictates the minimum pulse duration, as it assures a full Nyquist-sampling in the sampling space and conforms to amplitude and slew-rate limits.
For the present setup the total pulse duration was 7 ms. 
The time-steps are set to $10\,\mu\mathrm{s}$, resulting in 700 time bins, but in total $2\times 700$ bins to compute for the real and imaginary channels.

Each 2D RF pulse optimization was allowed 50 iterations with the midpoint method described in Ref.~\cite{vinding_optimal_2021-1}. 
Our workstation computed 25 pulses in parallel.
\subsection{Convolutional neural network}
\label{sec:cnn}
The presented deep neural network shown in Figure~\ref{FigMethodDeepControl} was inspired by the deep neural network of Ref.~\cite{vinding_deepcontrol_2021}.
The first image input layer imports the three maps of 64-by-64 pixels.
This is handed over to three 2D convolution layers with filter sizes 5, 3 and 2, and with 80, 600, and 60 filters in each, respectively. 
Each 2D convolution has a stride length of 2 in both directions, and each 2D convolution layer is followed by a Swish layer activation function.
After the three convolution layer there is a fully connected layer, which must have 1400 neurons, because it must match the output of the final regression layer, which is the 2D RF pulse (real and imaginary channels).
In between the fully connected layer and the regression layer, we have inserted the clipping layer, described in the next section, and marked in Figure~\ref{FigMethodDeepControl} by the red dashed lines.
\subsubsection{Clipping layer}
\label{sec:clip}
The clipping activation function layer serving as the main contribution in this work was made by modifying the function files of MATLAB's clipped Rectified Linear Unit (ReLU) layer for ease of implementation. 
For comparison, the ReLU layer function is 
\begin{equation}
f^\mathrm{ReLU}(x) = \max \left( 0,x\right) = \left\lbrace 
\begin{matrix} 
0,& x < 0\\
x,& x\geq 0
\end{matrix} \right.
\end{equation}
and the Swish layer function is
\begin{equation}
f^\mathrm{Swish}(x) =\frac{x}{1+e^{-x}}
\end{equation}
The clipped ReLU layer is:
\begin{equation}
f^\mathrm{clippedReLU}(x,c) = \min \left(\max \left( 0,x\right),c\right) = \left\lbrace 
\begin{matrix} 
0,& x < 0\\
x,& 0\leq x < c\\
c,& x\geq c
\end{matrix} 
\right.
\end{equation}
where $c$ is a clipping value.
And the modified, two-sided clipping version used herein to constrain RF amplitudes is
\begin{equation}
f^\mathrm{Clipping}(x,c) = \min \left(\max \left( -c,x\right),c\right) = \left\lbrace 
\begin{matrix} 
-c,& x < -c\\
x,& -c\leq x \leq c\\
c,& x > c
\end{matrix} 
\right.
\end{equation}
This type of clipping (or sometimes clamping) function was also used for clipping, e.g., deep neural network weights\cite{merolla_deep_2016}, and Adam\cite{kingma_adam_2017} optimization gradients\cite{shin_deep_2021}.

When computing the pulse library with optimal control, we imposed the hard constraint, $c'$ (Section~\ref{sec:oc}), on the RF amplitude (real and imaginary channels) through MATLAB's minimization function \textit{fmincon}, where upper and lower bounds are enforced strictly. 
An alternative to this is a weighted spill-out norm square penalty in the objective function, with a corresponding penalty gradient and Hessian\cite{goodwin_modified_2016}.

When the pulse library is imported into the deep learning framework, we normalize the RF waveforms by $c'$ such that they vary between $-1$ and $+1$ for numerical stability (similar as we did for the image inputs). 
Accordingly, the clipping value $c=1$. 

In the end, pulses predicted from the neural network are converted back to physical units through the $c'$ factor. 
\subsection{Deep learning}
\label{sec:dl}
We trained the deep neural network in Figure~\ref{FigMethodDeepControl} with the stochastic gradient descent with momentum algorithm\cite{rumelhart_learning_1986}. 
The momentum factor was 0.95.
The learn rate was constant at 0.006 and the $L_2$-regularization factor was $1\times10^{-4}$.
The total library of 500k cases was divided into fractions of $0.8:0.1:0.1$, which where used for training, validation and testing, resepectively.
The minibatch size was 1280 and we trained for 1000 epochs.
We used a Tesla P100 GPU (Nvidia, Santa Clara, USA) on the workstation for the deep learning.

For comparison, we also performed the same training on the neural network, where the clipping layer was deactivated.
Here, we call our network DeepControl, and we use the terms "clipped" or "non-clipped" to distinguish between the situations, where the clipping layer was activated or de-activated during training, respectively.

We stored the intermediate non-clipped networks for every epoch as checkpoints to monitor trends of pulse amplitude overshoots as deep learning improves the network.

\subsection{Realistic field maps}

While we perform our statistical analysis with the library test subset of 50k cases, we also tested each network (clipped and non-clipped) with a set of realistic field maps adopted from Ref.~\cite{grissom_ismrm_2016}, and shown by example in Figure~\ref{FigMethodDeepControl}.
Here, the target pattern was a 15-by-16 pixels handdrawn rectangle on a 64-by-64 pixels canvas in Paintbrush (Apple Inc., Cupertino, CA, USA). 
\section{Results}
\begin{figure}
\includegraphics[width=0.5\textwidth, angle=0]{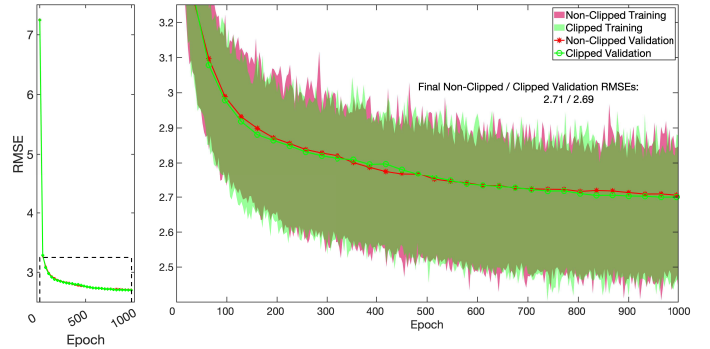}
\caption{Deep learning root-mean-square error (RMSE) as a function of epoch number for both the clipped and non-clipped DeepControl networks. Left, the RMSE of the validation subset. The dashed box signifies the zoomed view of the right plot, which also contains the RMSE of the training batches.}
\label{FigRMSE}
\end{figure}

\begin{figure}
\includegraphics[width=0.47\textwidth, angle=0]{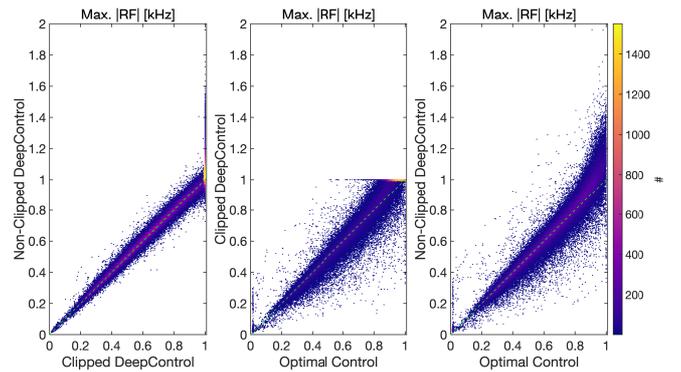}
\caption{Histograms of the maximum absolute real and imaginary RF amplitudes. Left, the non-clipped DeepControl network against the clipped DeepControl network. Center, the clipped DeepControl network against the optimal control cases. Right, the non-clipped DeepControl network against the optimal control cases. The green dashed lines are the desired trends. }
\label{FigMaxAmp}
\end{figure}

\begin{figure}
\includegraphics[width=0.47\textwidth, angle=0]{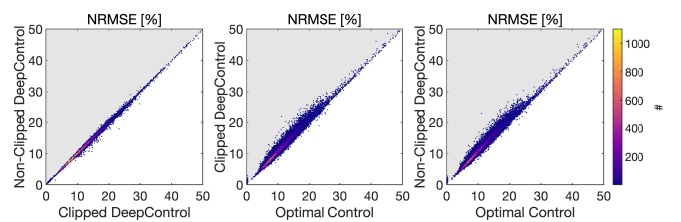}
\caption{Histograms of the normalized root-mean-square errors (NRMSE) of the magnetization components computed with Bloch simulations. Left, the non-clipped DeepControl network against the clipped DeepControl network. Center and right, the clipped and non-clipped DeepControl networks against the optimal control cases, respectively.}
\label{FigNRMSE}
\end{figure}

\begin{figure}
\includegraphics[width=0.47\textwidth, angle=0]{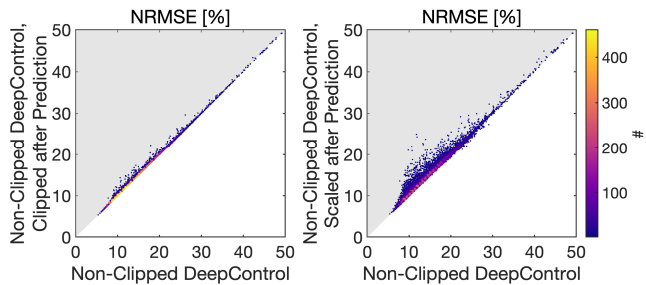}
\caption{NRMSE effects of regulating overshooting 2D RF pulses from the non-clipped DeepControl network, by clipping (left) or down-scaling (right) the 2D RF waveforms after pulse prediction. For clarity, the histograms only contain the overshooting pulses from the library test subset.}
\label{FigNRMSEclipscale}
\end{figure}

The deep learning progress of the clipped and non-clipped networks are shown in Figure~\ref{FigRMSE}.
Approximately one and a half to three days was spent for one thousand epochs. 
This conservative estimate is based on the fact that we ran multiple deep learning processes on the GPU simultaneously including various other demanding tasks occasionally.
The two training sessions (clipped and non-clipped) ended with near identical root-mean-square errors (RMSE) of the library validation subset.

The test subset of the supervision library was used to probe the network performances with unseen data.
With each network (clipped and non-clipped), we predicted 2D RF pulses from each test input, and contrasted these against each other and the corresponding optimal control pulses.

The maximum absolute RF values are contrasted in Figure~\ref{FigMaxAmp}. 
For these plots, we pooled the real and imaginary RF channels together. 
For example, ordinate counts above $1\,\mathrm{kHz}$ in the left and right plots correspond to either the real or imaginary or both RF channels of the non-clipped network pulses having values outside the range $[-c',+c']$.
If more than one instance of the RF channels spilled out, we only counted and plotted the maximum spill-out value. 
In total, 18.3\% of the non-clipped network pulses overshot the $c' = 1\,\mathrm{kHz}$ limit somehow.
The amount of overshoot was $84\pm78\,\mathrm{Hz}$ (mean $\pm$ standard deviation in the test subset). 
From the non-clipped networks stored at each epoch, we have estimated the amount of overshoot to be $17.6 \pm 1.2 \%$.
Figure S1 in the Supplementary Material shows the test case overshoot percentage as a function of epochs together with a histogram.
The right panel of Figure~\ref{FigMaxAmp} shows that the non-clipped network for amplitudes lower than $0.75\,\mathrm{kHz}$ follows optimal control cases quite well. 
Above $0.75\,\mathrm{kHz}$ and beyond $1\,\mathrm{kHz}$ the non-clipped network becomes more diffuse.
The clipped network (center panel of Figure~\ref{FigMaxAmp}) as expected piles up just below the $1\,\mathrm{kHz}$ limit.
The two clipped and non-clipped networks (left panel of Figure~\ref{FigMaxAmp}) are quite similar, until around the $1\,\mathrm{kHz}$ limit.

The pulses were simulated in Bloch simulations and we measured the normalized root mean square errors (NRMSE) of the resulting excitation profiles with respect to the target excitation profiles, as shown in Figure~\ref{FigNRMSE}.
In conventional pulse optimization the NRMSE or a similar merit of the magnetization profile constitutes a direct measure of the optimization performance.
In this framework, the NRMSE of the magnetization profile is indirect, because the magnetization profile is not part of the deep learning objective.
The deep learning measures the RMSE of the network-predicted pulses with respect to the library pulses.
However, Figure~\ref{FigNRMSE} (center and right panels) show a good correspondence between the DeepControl and optimal control magnetization responses, and the (left) panel shows a near identical performance between the two clipped and non-clipped networks.

Although the clipped network solves the problem of pulse overshoots, we investigated the effects of regulating pulses from the non-clipped network that overshot the limit through two options as proposed in Ref.~\cite{vinding_ultrafast_2019}. 
One option is to clip the pulse manually.
If the internal checks of the MRI system does not block the pulse, the RF transmit chain will likely perform the clipping at some point.
Another option is to down-scale the entire pulse waveform.
The effects of performing clipping and scaling are shown in Figure~\ref{FigNRMSEclipscale}.
Clipping and scaling after pulse prediction increased the NRMSE in the majority of the cases.
Scaling had the greatest (negative) effect.

On the workstation, the pulse prediction times (mean $\pm$ standard deviation in the test subset) for the clipped and non-clipped networks were $5.8\pm1.3\,\mathrm{ms}$ and $3.9\pm0.3\,\mathrm{ms}$, respectively.
The corresponding optimal control pulse computation times were around 20 to 30 seconds.

The video in the Supplementary Material displays a random selection of 10k simulations from the library test subset.
It shows the photograph-based $B_0$ and $B^+_1$ field maps, and the randomly generated target areas, i.e., from the library input section.
It also shows the 2D RF pulses predicted by the clipped and non-clipped networks, and the corresponding optimal control pulses, i.e., the latter from the library output section.
Finally, the simulated magnetization responses are shown.

\subsection{Realistic field maps}
\begin{figure}
\includegraphics[width=0.47\textwidth, angle=0]{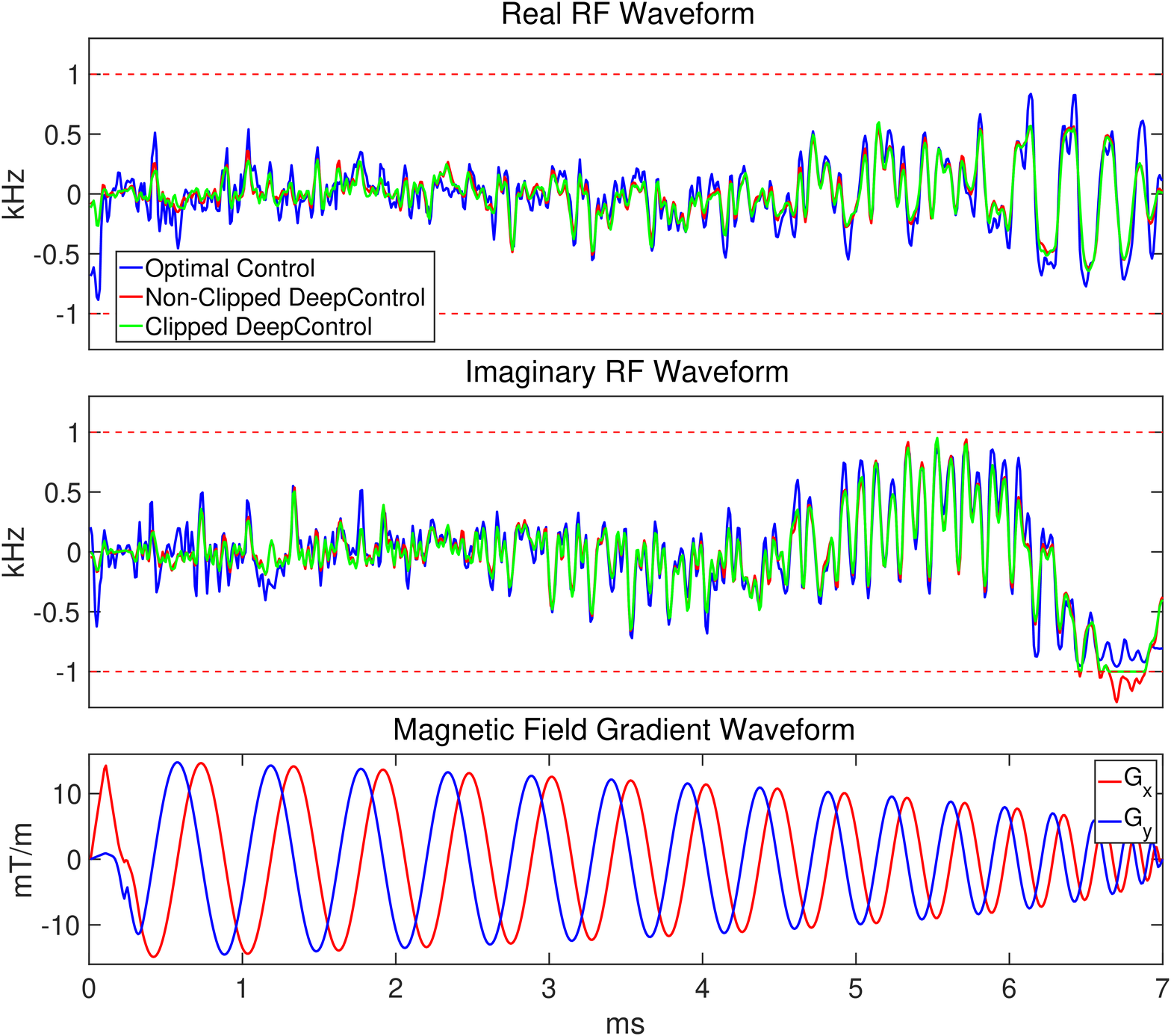}
\caption{The 2D RF pulses computed with realistic field maps, see Figure~\ref{FigMagn}. Top/central panel, the real/imaginary RF waveforms: optimal control (blue), non-clipped DeepControl (red) and clipped DeepControl (green). The bottom panel shows the magnetic field gradient waveform accompanying the RF pulses in experiments for spatial encoding.}
\label{FigPulses}
\end{figure}

\begin{figure}
\includegraphics[width=0.47\textwidth, angle=0]{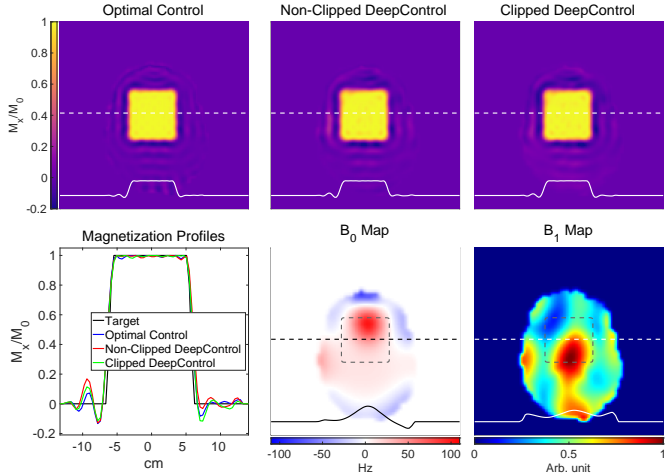}
\caption{Excitation maps (upper row) of the optimal control (left), non-clipped DeepControl (center) and clipped DeepControl (right) pulse responses. The dashed lines show, where the profiles (full lines) are from. They are overlaid in the lower left panel. The lower center and right panels show the $B_0$ and $B^+_1$ field maps, including the profiles and also the dashed frames signifying the target area.}
\label{FigMagn}
\end{figure}

The DeepControl pulses predicted for the realistic field maps are shown together with the optimal control pulse in Figure~\ref{FigPulses}. 
We observe a pulse overshoot of 255 Hz of the imaginary RF waveform for the non-clipped pulse.
We found (not shown) that this pulse overshoot vanished if the target mask was trimmed to 13-by-14 pixels instead. 
The clipped network is as shown not overshooting the limit.

The bottom panel of Figure~\ref{FigPulses} shows the magnetic field gradient waveform that in simulations and in the optimal control framework is underlying all 2D RF pulses, i.e., it is played simultaneous to the RF pulses.
While the magnetic field gradient waveform could have been further optimized in the optimal control and correspondingly been subject to a neural network output, we chose not to pursue this because this particular oscillating waveform is already time-optimized\cite{lustig_fast_2008}.

\begin{figure}
\includegraphics[width=0.47\textwidth, angle=0]{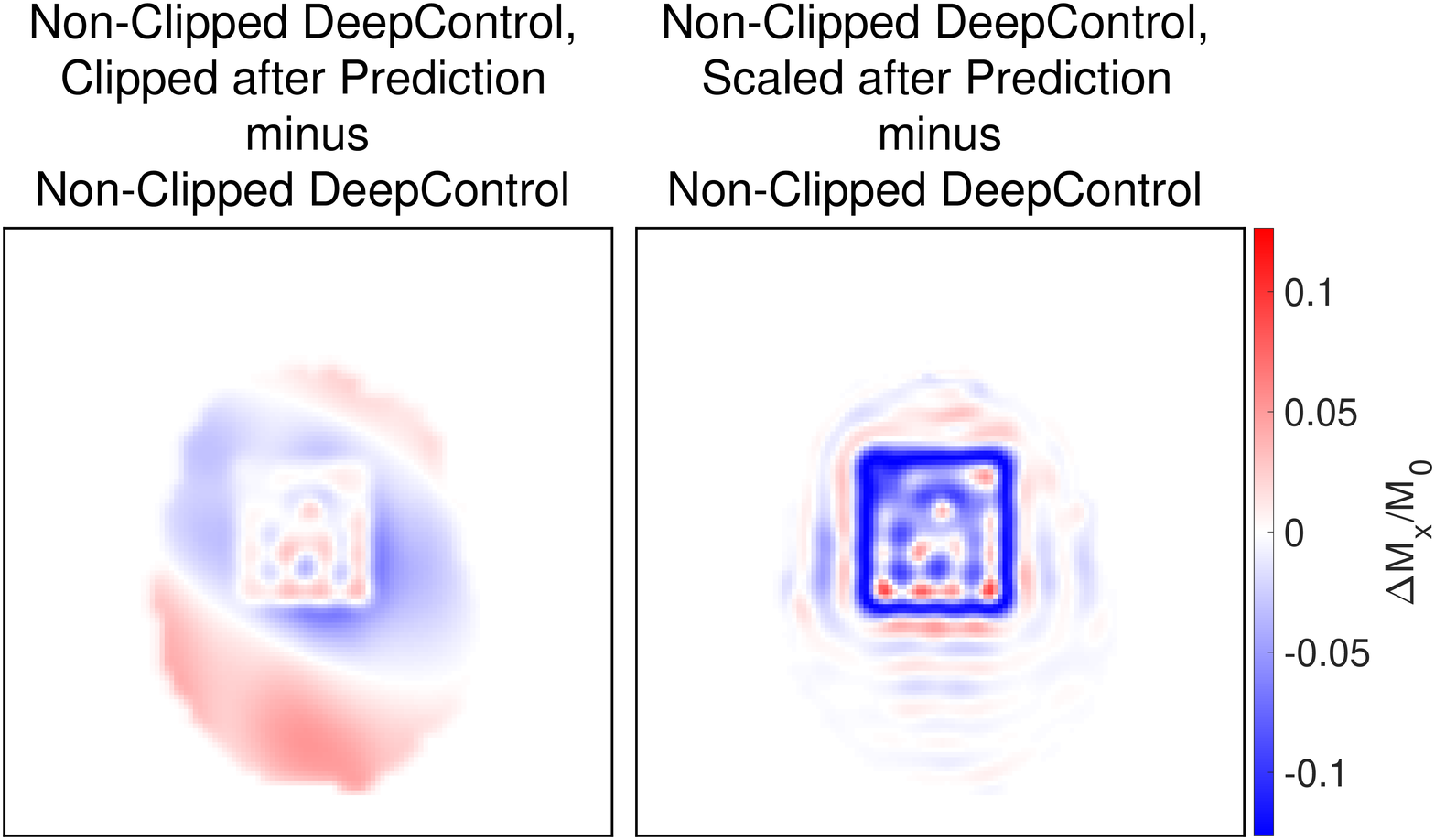}
\caption{Effects on the magnetization patterns of regulating the amplitude overshoot of the non-clipped DeepControl pulse from Figure~\ref{FigPulses}. The pulse has been clipped (left) or down-scaled (right) after the non-clipped DeepControl network predicted the pulse.}
\label{FigMagnDiff}
\end{figure}

The pulses of Figure~\ref{FigPulses} return the magnetization profiles shown in of Figure~\ref{FigMagn} also displaying the realistic field maps.  
We see a near-identical performance between the three contestants.
However, regulating the pulse amplitude by clipping or scaling, as discussed in the previous section, influenced the magnetization patterns as shown in Figure~\ref{FigMagnDiff}.

\section{Discussion}

A clipping layer has been introduced to the DeepControl neural network, that effectively prevents pulse amplitude overshoots.
As compared to the same network except with the clipping layer deactivated, there is no significant difference in performance.
Both networks (clipped and non-clipped) also yield almost the same performance as the conventional method of optimal control.
The gain is a speed gain in pulse computation time by a factor of $>3000$.

While the communication has mainly concerned prevention of RF amplitude overshoots internally in the clipped neural network, we also investigated how regulating RF amplitude overshoots of a non-clipped network affected the magnetization responses, by simply clipping or scaling the overshooting pulses manually, as proposed in Ref.~\cite{vinding_ultrafast_2019}.
Our non-clipped network produced overshooting pulses in 18.3\% of test cases.
But for the limit of 1 kHz, the amount of overshoot was one average below 0.1 kHz. 
For the type of magnetic field gradient waveform as shown in Figure~\ref{FigPulses}, the accompanying 2D RF pulse waveforms often appear spiky, with high transients, but the highest amplitudes typically happen near the end, where the bulk magnetization is flipped.
Therefore, manually clipping such 2D RF pulses does not, for the majority of our testcases, yield huge increases in NRMSE. 
Only a small fraction of the pulse waveform is clipped.
But still, it may lead to changes of the magnetization profiles of around 10\% as shown in one example.
Down-scaling pulses to fit within the boundary, a common way to change the desired FA, lead to higher NRMSE increases, mainly because we change the entire pulse waveform.
As also suggested\cite{vinding_ultrafast_2019}, we could also remove pulse overshoots, by trimming the target pattern area in our realistic case study. 
This approach can obviously be inconvenient, as there is no guarantee for success, or the final target may be unsatisfactory.
To minimize the risk of pulse overshoot, one might further constrict the conventional pulses, in our case within the optimal control, e.g., setting the limit to 0.9 kHz\cite{vinding_ultrafast_2019}.
This will likely affect the overall performance of both the conventional optimal control and the DeepControl pulses, but it also alleviates the hardware.
The clipped DeepControl network solves the overshooting problem and enables a more efficient use of the system.

In Ref.~\cite{vinding_deepcontrol_2021} we used ReLU layers after convolution layers. 
In this study, we have replaced the ReLU layers with Swish layers.
This trades substantially longer training for lower RMSEs.
But for faster progress, it is possible to obtain good results with ReLU layers (data not shown). 
We adjusted many of the other network and deep learning parameters before switching from ReLU to Swish layers .
For example, the number of filters in the convolution layers, the mini-batch size, the learn rate and momentum factor.
Since sweeping the vast multi-dimensional parameter space is a considerable task given the long deep learning time (with either ReLU or Swish layers), our success criterion was satisfactory NRMSE values and visual quality inspection of the magnetization profiles from simulations, which is in line with conventional pulse optimization approaches.

Average power did not raise concern in the previous study\cite{vinding_deepcontrol_2021}, but pulse optimization pipelines sometimes include average power limits, and/or even local/global SAR constraints.
We hope to address these constraints in future studies, especially local SAR for multi-channel RF systems\cite{vinding_local_2017}.
\section{Conclusion}
Utilizing deep learning in magnetic resonance imaging, to improve image reconstruction or diagnoses, or facilitate novel control designs receives an increasing interest these years.
Here we have shown that a custom-made clipping layer in a deep neural network actively limits predicted controls, such that they stay within user prescribed boundaries or hardware restrictions.
This sort of control limit was in the previous networks only enforced in the supervision library. 
Hence, the deep learned neural network was not guaranteed to adopt these limits.
Here we have shown that the clipping layer does not interfere with training progress nor the statistical performance of the controls.
\section*{Acknowledgments}
We thank VILLUM Fonden, Eva og Henry Fraenkels Mindefond, Harboefonden, and Kong Christian Den Tiendes Fond. 
We also thank C.S Aigner, I. Kuprov, D.F. Hansen, I. Maximov, and D.L. Goodwin for fruitful discussions.
\section*{Code Availability}
Our repository at https://github.com/madssakre/DeepControl3 will contain the proposed method.
\bibliographystyle{elsarticle-num-names}
\bibliography{DCCBib}

\end{document}